\documentclass[aps,pr,twocolumn,showpacs,showkeys,superscriptaddress,a4paper]{revtex4}

\usepackage{graphicx,amssymb,amsmath}
\usepackage{dcolumn}
\usepackage{color}

\begin{document}
\bibliographystyle{revtex}
%

\title{Upper limits  for   a narrow resonance  in the reaction
$p + p \to K^+ + (\Lambda p)$}

\affiliation{CSSM, School of Chemistry and Physics,
University of Adelaide, Adelaide SA 5005, Australia}

\affiliation{Helmholtz-Institut f\"{u}r Strahlen- und Kernphysik der Universit\"{a}t Bonn, 53115 Bonn, Germany}

\affiliation{Laboratory for High Energies, JINR Dubna, Russia}

\affiliation{Fachbereich Physik, Universit\"{a}t Duisburg-Essen,
Duisburg, Germany}

\affiliation{Institut f\"{u}r Kernphysik, Forschungszentrum J\"{u}lich,
J\"{u}lich, Germany}

\affiliation{J\"{u}lich Center for Hadron Physics, Forschungszentrum J\"{u}lich,
J\"{u}lich, Germany}

\affiliation{Technical University, Kosice, Kosice, Slovakia}

\affiliation{P.~J.~Safarik University, Kosice, Slovakia}

\affiliation{Institute of Nuclear Physics, PAN, Krakow, Poland}

\affiliation{Institute of Physics, Jagellonian University, Krakow, Poland}

\affiliation{Nuclear Physics Division, BARC, Mumbai, India}

\affiliation{Physics Faculty, University of Sofia, Sofia, Bulgaria}

\affiliation{Physikalisches Institut. Universit\"{a}t T\"{u}bingen, Germany}

\author{A.~Budzanowski\footnote{deceased}}
\affiliation{Institute of Nuclear Physics, PAN, Krakow, Poland}

\author{A.~Chatterjee}
\affiliation{Nuclear Physics Division, BARC, Mumbai, India}

\author{H.~Clement}
\affiliation{Physikalisches Institut. Universit\"{a}t T\"{u}bingen, Germany}


\author{P.~Hawranek}
\affiliation{Institute of Physics, Jagellonian University, Krakow, Poland}

\author{F.~Hinterberger}
\email[corresponding author Frank Hinterberger: ]{fh@hiskp.uni-bonn.de}
\affiliation{Helmholtz-Institut f\"{u}r Strahlen- und Kernphysik der
Universit\"{a}t Bonn, 53115 Bonn, Germany}

\author{R.~Jahn}
\affiliation{Helmholtz-Institut f\"{u}r Strahlen- und Kernphysik der Universit\"{a}t Bonn, 53115 Bonn, Germany}

\author{R.~Joosten}
\affiliation{Helmholtz-Institut f\"{u}r Strahlen- und Kernphysik der
Universit\"{a}t Bonn, 53115 Bonn, Germany}

\author{K.~Kilian}
\affiliation{Institut f\"{u}r Kernphysik, Forschungszentrum J\"{u}lich,
J\"{u}lich, Germany}
\affiliation{J\"{u}lich Center for Hadron Physics, Forschungszentrum J\"{u}lich,
J\"{u}lich, Germany}

\author{Da.~Kirillov}
\affiliation{Fachbereich Physik, Universit\"{a}t
Duisburg-Essen, Duisburg, Germany}
\affiliation{Institut f\"{u}r Kernphysik, Forschungszentrum J\"{u}lich,
J\"{u}lich, Germany}

\author{Di.~Kirillov}
\affiliation{Laboratory for High Energies, JINR Dubna, Russia}

\author{S.~Kliczewski}
\affiliation{Institute of Nuclear Physics, PAN, Krakow, Poland}

\author{D.~Kolev}
\affiliation{Physics Faculty, University of Sofia, Sofia, Bulgaria}

\author{M.~Kravcikova}
\affiliation{Technical University, Kosice, Kosice, Slovakia}

\author{M.~Lesiak}
\affiliation{Institute of Physics, Jagellonian
University, Krakow, Poland}

\author{H.~Machner}
\affiliation{Fachbereich Physik, Universit\"{a}t Duisburg-Essen,
Duisburg, Germany}

\author{A.~Magiera}
\affiliation{Institute of Physics, Jagellonian University, Krakow, Poland}

\author{G.~Martinska}
\affiliation{P.~J.~Safarik University, Kosice, Slovakia}

\author{N.~Piskunov}
\affiliation{Laboratory for High Energies, JINR Dubna, Russia}

\author{D.~Proti\'c}
\affiliation{Institut f\"{u}r Kernphysik, Forschungszentrum J\"{u}lich,
J\"{u}lich, Germany}

\author{J.~Ritman}
\affiliation{Institut f\"{u}r Kernphysik, Forschungszentrum J\"{u}lich,
J\"{u}lich, Germany}
\affiliation{J\"{u}lich Center for Hadron Physics, Forschungszentrum J\"{u}lich,
J\"{u}lich, Germany}

\author{P.~von Rossen}
\affiliation{Institut f\"{u}r Kernphysik, Forschungszentrum J\"{u}lich,
J\"{u}lich, Germany}
\affiliation{J\"{u}lich Center for Hadron Physics, Forschungszentrum J\"{u}lich,
J\"{u}lich, Germany}

\author{J.~Roy}
\affiliation{Nuclear Physics Division, BARC, Mumbai, India}

\author{A.~Sibirtsev}
\affiliation{CSSM, School of Chemistry and Physics,
University of Adelaide, Adelaide SA 5005, Australia}

\author{I.~Sitnik}
\affiliation{Laboratory for High Energies, JINR Dubna, Russia}

\author{R.~Siudak}
\affiliation{Institute of Nuclear Physics, PAN, Krakow, Poland}

\author{R.~Tsenov}
\affiliation{Physics Faculty, University of Sofia, Sofia, Bulgaria}

\author{K.~Ulbrich}
\affiliation{Helmholtz-Institut f\"{u}r Strahlen- und Kernphysik der Universit\"{a}t Bonn, 53115 Bonn, Germany}

\author{J.~Urban}
\affiliation{P.~J.~Safarik University, Kosice, Slovakia}

\author{G.~J.~Wagner}
\affiliation{Physikalisches Institut. Universit\"{a}t T\"{u}bingen, Germany}

\collaboration{The HIRES Collaboration}




\date{\today}

\begin{abstract}%
The reaction $pp\to K^+ + (\Lambda p)$
has been  measured at $T_p=1.953$~GeV and $\Theta=0^{\circ}$
with a high missing-mass resolution
in order to study the $\Lambda p$ final state interaction.
Narrow $S=-1$ resonances predicted by bag model calculations
are not visible in the missing-mass spectrum.
Small structures observed in a previous experiment
are not confirmed.
Upper limits for the production cross section of a narrow resonance
are deduced for missing-masses between 2058 and 2105 MeV.
\end{abstract}

\maketitle

\section{Introduction}

A high-resolution study of the reaction $p+p\to K^+ + (\Lambda p)$
has been performed by the HIRES Collaboration \cite{bud10,bud10a} using the proton beam
of the Cooler Synchrotron COSY \cite{mai97} and the magnetic spectrograph
BIG KARL \cite{dro98,bojo02} at the Research Center  J\"{u}lich.
The aim of the experiment was to study the $\Lambda {p}$
final state interaction (FSI) and to search for
narrow strangeness $S=-1$ resonances.
Concerning the FSI, first results have been published \cite{bud10}.
The present paper deals with the search for
a narrow strangeness $S=-1$ resonance.
We use only the missing-mass data measured at 1.953~GeV beam energy \cite{bud10} and
ignore a small portion of data below the $\Sigma N$ threshold taken at some higher beam energy \cite{bud10a}.

Predictions of strange dibaryons are summarized in  a recent review by Gal \cite{gal10}.
A confined six-quark  state $(Q^6)_1$ with $S=-2$, the so called H-dibaryon (from Hexaquark), 
has been predicted to be 
the lowest-lying dibaryon state \cite{jaf77}.
It has been of prime interest both theoretically and experimentally.
In the strangeness -1 sector narrow dibaryon resonances $D_s$ and $D_t$ are predicted to be located at
about 55 MeV and 95 MeV above the  $\Lambda p$ threshold \cite{aer84,aer85,mul79,mul80}.
The width of the lowest resonance $D_s$ is estimated to be less
than about 1~MeV.


Experimentally, the inclusive reaction $pp\to K^+ + (\Lambda p)$
was first studied  with a rather low missing-mass resolution
\cite{fhint:mel65,fhint:ree68,fhint:hog68}.
There were also exclusive measurements of the reaction
${p}{p}\to {K^+}{\Lambda}{p}$ by the COSY-TOF Collaboration \cite{bil98,abd06,abd10}
and total cross section measurements by the 
COSY-11  \cite{bal98} and 
COSY-ANKE Collaboration \cite{val07,val10}.
The first high-resolution measurement of $pp\to K^+ + (\Lambda p)$
has been performed at SATURNE II  
with proton beam energies of 2.3 and 2.7 GeV
using the
spectrometer SPES4 \cite{fhint:sie94}.
The outgoing kaons were detected at forward angles with a
high momentum resolution in the focal plane of the SPES4 spectrometer.
The missing-mass spectra show characteristic enhancements near
the $\Lambda$p and $\Sigma$N thresholds.
At 2.3~GeV and $\Theta=10^{\circ}$,
a sharp peak  has been observed
in the missing-mass spectrum at $2096.5\pm 1.5$~MeV
above the background which is due to the $\Lambda $p continuum.
The peak  amounts to about 3.5 standard deviations.
A similar peak  corresponding to a missing-mass of $2098.0\pm 1.5$~MeV
has been observed at 2.7~GeV beam energy and $12.6^{\circ}$.
The statistical accuracies were not
high enough to exclude an accidental statistical fluctuation,
however, the peak has been observed twice under
different experimental and kinematical conditions
and the peak energies coincide within the experimental error.
It is also interesting to note that
the observed peaks were located near the predicted mass
of the lowest S = -1 dibaryon $D_s$ \cite{aer84,aer85}.
Therefore,
the reaction
$pp\to K^+ + (\Lambda p)$ has been studied
with an especially  high accuracy in the
missing-mass range between
2090 and 2110~MeV.

In Sec. II we give a short description of the experiment.
In Sec. III  we sketch the predictions of narrow dibaryon resonances.
The effect of the lowest dibaryon $D_s$ in the total cross section
of the free $\Lambda {p} \to \Lambda {p}$ scattering is discussed in Sec. IV.
The formalism to describe a narrow resonance embedded in the continuum
of the reaction $pp\to K^+ + (\Lambda p)$ is presented in Sec. V.
In Sec. VI we deduce upper limits for the production cross section
of the predicted resonance $D_s$. Conclusion and discussion
follow in Sec. VII.

\section{Experiment}\label{experiment}

\begin{figure}[t!]
\begin{center}
\includegraphics[width=0.50\textwidth]{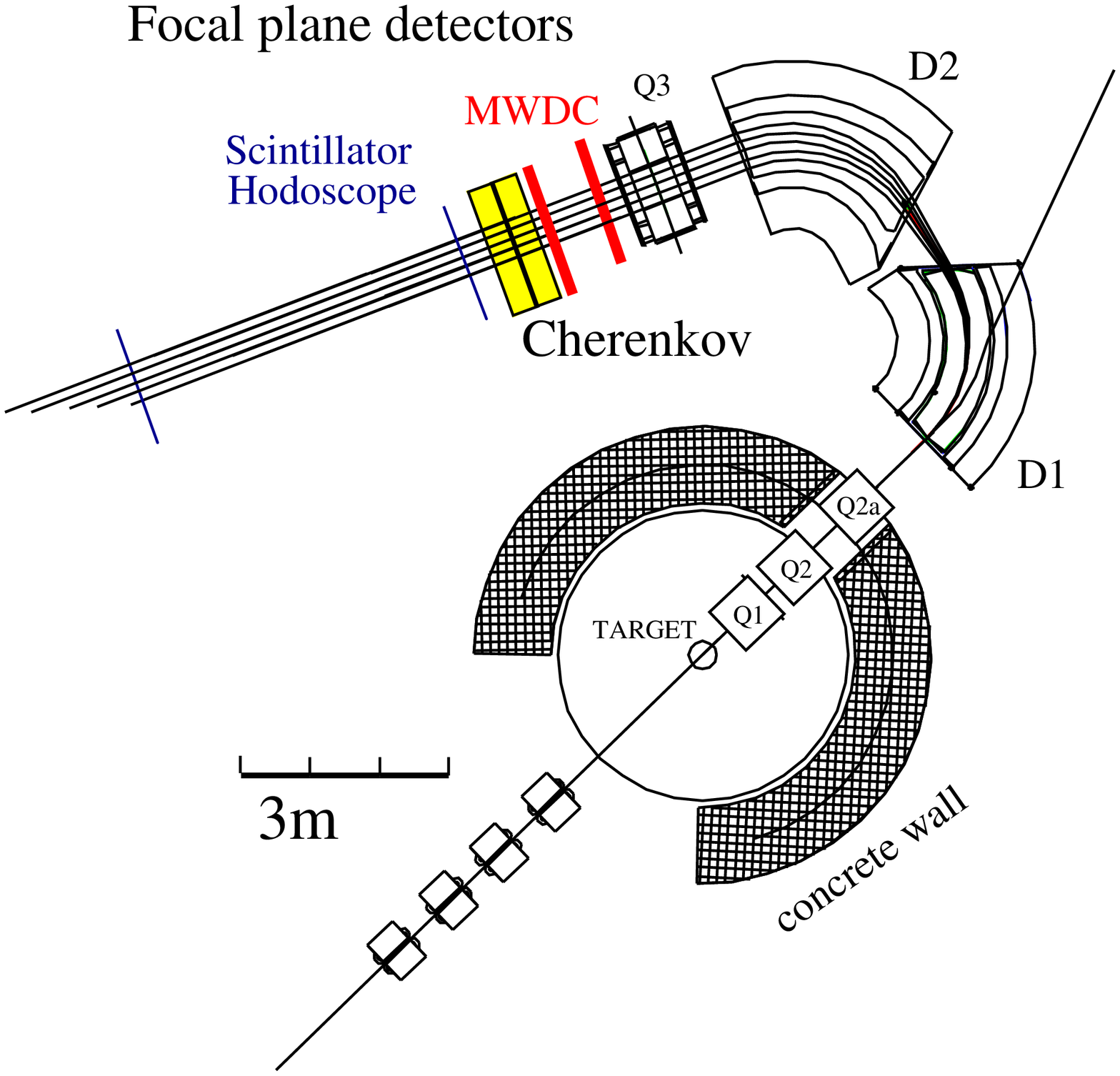}
\includegraphics[width=0.39\textwidth]{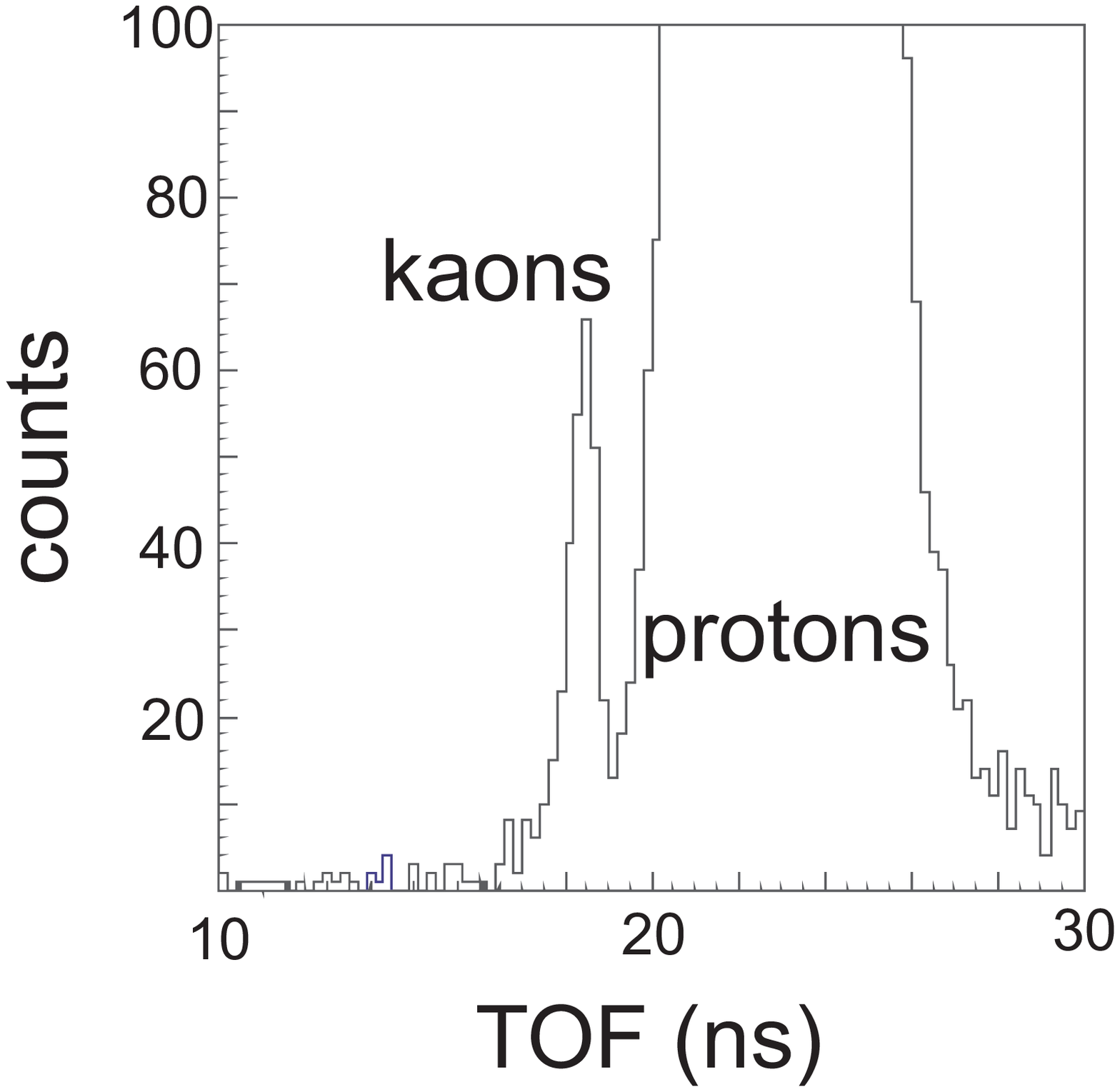}
\end{center}
\caption{Top: Layout of the magnetic spectrograph BIG KARL.
The charged particle tracks are detected in the focal
plane using  two stacks of multiwire drift chambers,
two threshold Cherenkov detectors, and 
two scintillator hodoscopes. Bottom: TOF spectrum with pion suppression
by two Cherenkov detectors.}
\label{bigkarl}
\end{figure}

Here, we give a short description of the experiment
which was already reported in \cite{bud10}.
The reaction   $p + p \to K^+ + (\Lambda p)$
was measured at $0^{\circ}$
using the proton beam from the  cooler synchrotron COSY,
the  magnetic spectrograph BIG KARL \cite{bojo02}
and a 1.0~cm thick liquid hydrogen target (see Fig.~\ref{bigkarl}).
The momentum of the incoming proton beam
was 2.735~GeV/c corresponding to a kinetic energy of 1.953~GeV.
The scattered  particles in the
momentum range  930 - 1110~MeV/$c$
were detected in the focal
plane using  two stacks of multiwire drift chambers,
two threshold Cherenkov detectors \cite{siu08},  and 
two scintillator hodoscopes.
The absolute beam momentum was found 
by measuring simultaneously the tracks of $K^+$ particles and deuterons from the
reactions $p + p \to K^+ + (\Lambda p)$ and $p + p \to d + \pi^+$
at fixed BIG KARL momentum 1070~MeV/$c$
and fitting the kinematic parameters.
The absolute precision of the beam
momentum was 0.15~MeV/$c$. 
The ratio of beam momentum to scattered particle momentum
allowed for a measurement at $0^{\circ}$.
In the first dipole magnet of BIG KARL the beam was
magnetically separated from the scattered particles
and guided through the side exit of the outer yoke
into the beam dump. 
Thus, the huge background from dumping the
beam within the spectrometer was avoided.
Particle identification was performed using the energy loss ($\Delta E$) and time of flight (TOF)
information from the scintillator hodoscopes.
In addition two threshold Cherenkov detectors \cite{siu08} were used in order to 
achieve a very high pion suppression factor of $10^5$.
The momentum of the kaon was measured and the missing-mass of the $\Lambda p$ system was deduced. 
In order to cover the missing-mass range 2050 -- 2110~MeV 
the data were taken using
three overlapping settings of the spectrograph 
(mean momenta: 1070, 1010 and 960~MeV/$c$).
The incoming beam was not changed during those
measurements. 
The relative precision of the momenta  of the three settings 
was 0.1 MeV/$c$.

\begin{figure}[h]
\begin{center}
\includegraphics[width=0.49\textwidth]{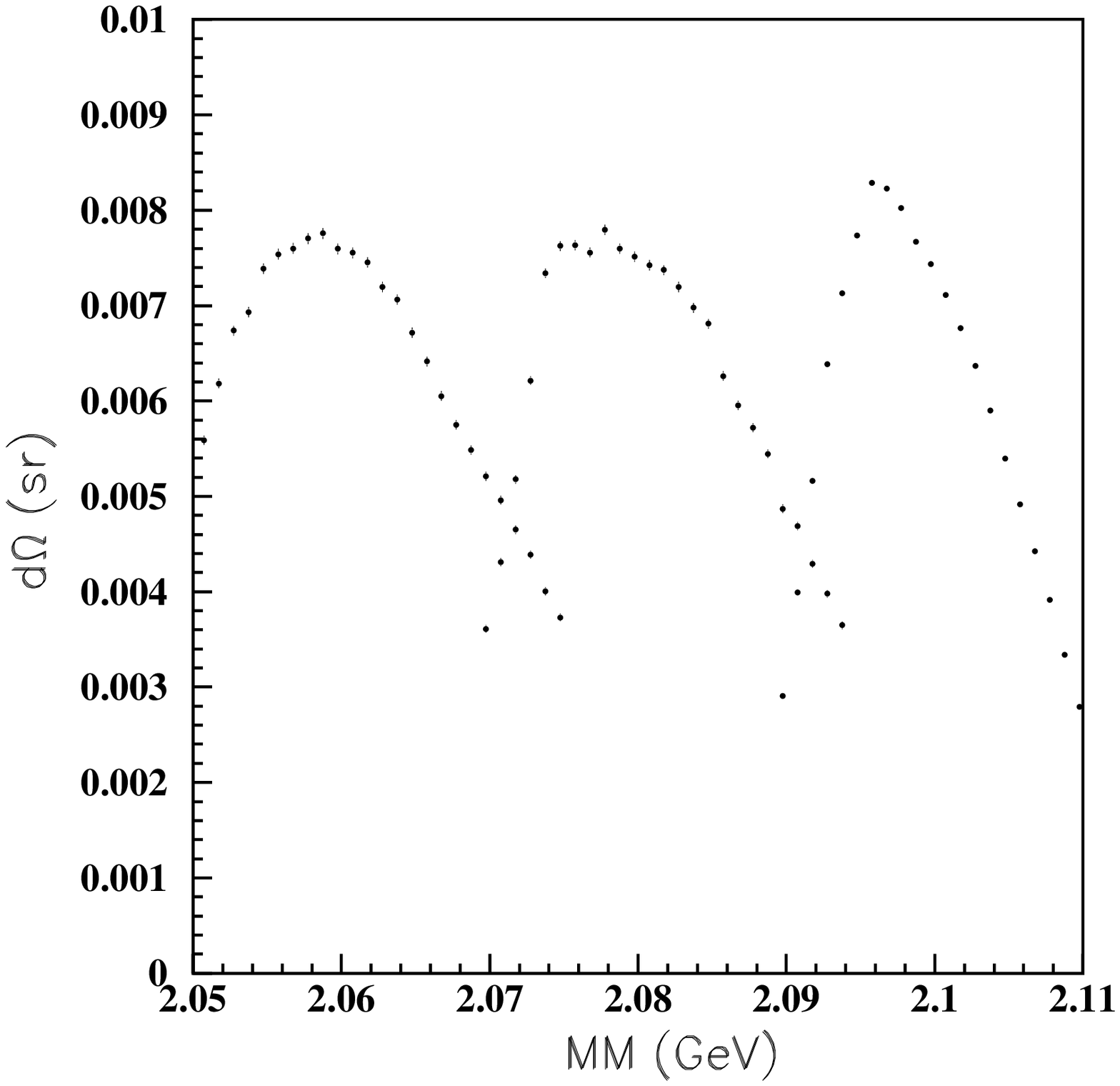}
\includegraphics[width=0.49\textwidth]{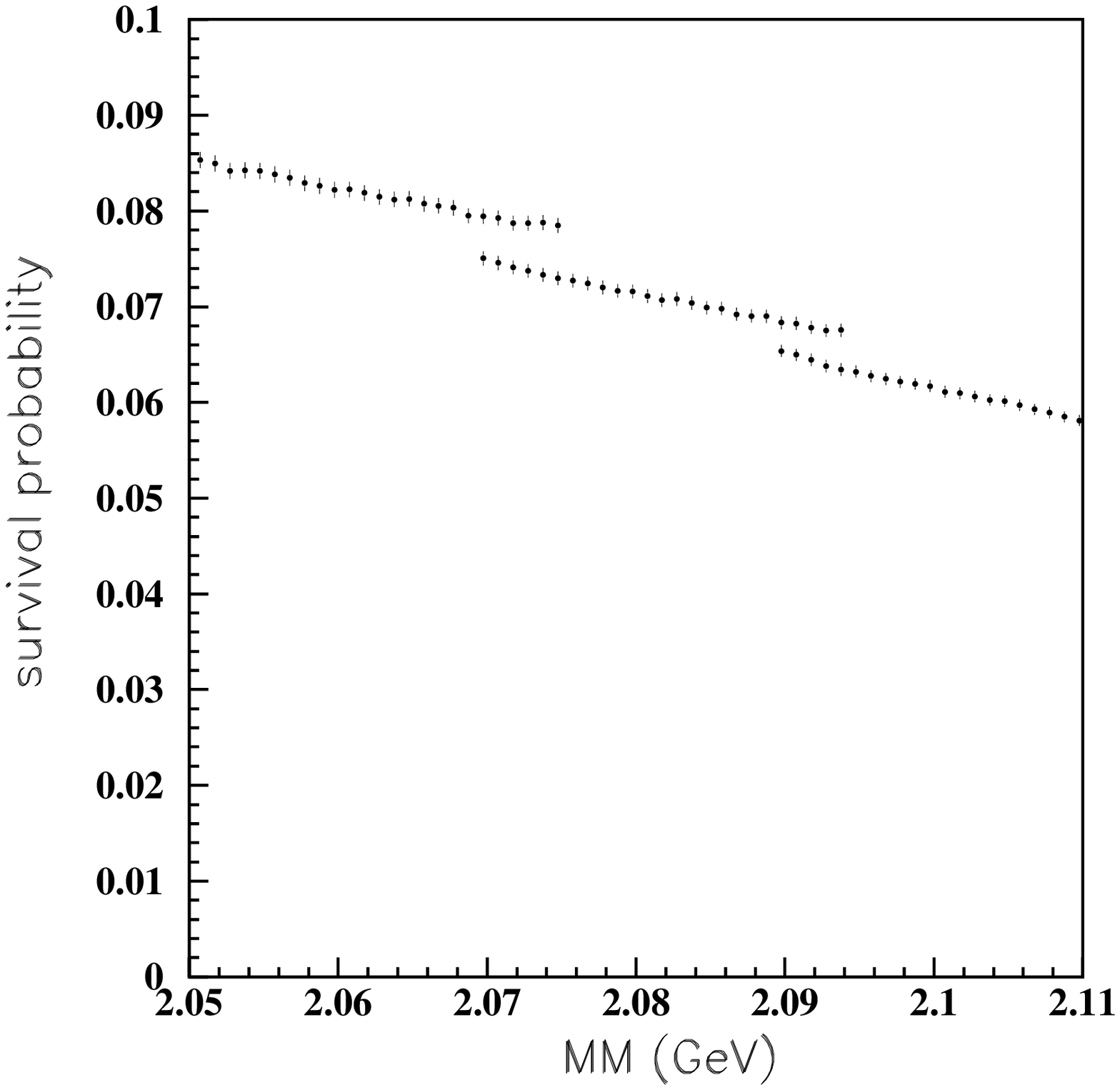}
\caption{Top: Acceptance correction functions. Bottom:
Kaon survival probability} 
\label{acceptance}
\end{center}
\end{figure}

Acceptance corrections with respect to solid angle and momentum were taken from
Monte Carlo calculations. 
The acceptance correction functions  contained  the magnetic spectrograph
momentum acceptance
around   $0^{\circ}$  
with cuts on the horizontal emission angle $\Theta_x=\arctan(p_x/p_z)$ and the
vertical emission angle $\Theta_y=\arctan(p_y/p_z)$ with $p_x$ and $p_y$ transversal
momentum components and $p_z$ the longitudinal component.
The cuts  for a given missing-mass bin 
corresponded to the measured solid angle $d\Omega$. They contained also  
the detector efficiency corrections.
 The detector efficiency  included efficiency  of scintillator detectors which
 determined trigger and particle identification and
magnetic spectrograph efficiency with field inhomogeneity at the edges of the
acceptance.
Acceptance corrections as determined by Monte Carlo simulations were checked
by use of
experimental distributions of simultaneously 
measured  pions.
They are shown in Fig.~\ref{acceptance}.
The acceptance correction function of the 960~MeV/$c$ setting looks
different due to a slightly different 
$\Theta_x$ and $\Theta_y$ 
cut.

The kaon decay along flight path was taken into account for each individual
trajectory.
Path lengths were obtained from  
calculations of tracks in the magnetic field
with the Turtle code \cite{Turtle}. They  were checked by experimentally deduced
values from time of
flight measurements of pions and protons.
The cross section error due to survival probability is less than 1~\%. 
The final kaon survival probability 
averaged for a given missing-mass bin
is presented in the bottom panel of Fig.~\ref{acceptance}.

The relative normalizations of the three different spectrograph settings were deduced from
luminosity monitors located in the target area
which were independent of the spectrograph settings.
The relative normalization errors 
due to the luminosity measurement were negligibly small. 
The relative normalization errors due to the acceptance
corrections  were estimated to be less than  2~\%.

The absolute cross section normalization was determined by measuring the 
luminosity  as described in \cite{bojo02}.
At the beginning of each beam period we calibrated two  luminosity monitors 
counting the left and right scattered particles 
from the target as function of the number of beam particles.
To this end, the beam current was highly reduced in order to
count the beam particles individually with a fast scintillator hodoscope in the beam dump.
For the calibration, the dependence of the luminosity signal on the beam intensity
was fitted by a linear function.
The resulting beam intensity error amounted to about 5~\%. The density of the 
bubble-free liquid hydrogen target  ($\rho=0.0775$~g/cm$^3$)
with 1~$\mu$m thick  mylar foil windows was kept constant by stabilizing the
temperature to $15.0\pm 0.5$~K using
a high-precision temperature control  \cite{kilian02}.

The target thickness, i.e. the distance between entrance and exit window (nominal 1~cm) 
was precisely measured with a calibrated optical telescope.
The target thickness error was about 5~\%. 
The overall systematic normalization error was estimated to be  10~\%.
The missing-mass spectrum is shown in Fig.~\ref{spectrum}.

In view of the predicted narrow dibaryon resonances in the $\Lambda{p}$ system
a high missing-mass resolution was required.
The missing-mass resolution depends on the spread of the beam momentum, 
the beam spot size at the target, the target thickness, the momentum
resolution of the magnetic spectrograph and the 1~MeV bin width.
The effective resolution function $f(M,M')$ has been deduced 
in \cite{bud10} 
by a least-square fit to the
sharply rising
double differential cross section
${\rm d}^2\sigma/({\rm d}\Omega_{K}{\rm d}M_{\Lambda{p}})$
near the $(\Lambda {p})$ threshold, see Fig.~\ref{spectrum}.
It can be represented by a Gaussian density distribution
with a standard deviation of $\sigma_M=0.84$ MeV.
It should be noted that
the width $\sigma_M$ depends mainly on
the spread of the beam momentum and the 1~MeV bin width.
The contributions from  the target thickness, the beam spot size at the target (3 mm) and the 
momentum resolution of the magnetic spectrograph are
negligibly small. Since the COSY beam has not been changed during
the measurements the resolution function $f(M,M')$ is constant  
in the full range of missing-masses.

\section{Predictions of narrow dibaryon resonances in the $\Lambda p$ system}

Concerning the predictions of narrow $S=-1$ dibaryon resonances in the
$\Lambda {p}$ system
we refer to calculations  of Aerts and Dover \cite{aer84,aer85}.
They studied the decay width of strangeness $S=-1$ six-quark bag states
predicted by the Nijmegen group \cite{mul79,mul80} on the basis of the MIT bag model.
The predicted $S=-2$ H-dibaryon with quark structure $(Q^6)_1$ is the strongest bound dibaryon state \cite{jaf77}.
In case of $S=-1$ states the color-magnetic force is less strong and 
hence the lowest $(Q^6)_1$  state is predicted to be about 120 MeV above the $\Lambda N$ threshold \cite{mul80}. 

The lowest-lying $S=-1$ resonances exhibit a quark cluster structure 
$Q^4\bigotimes Q^2$. The singlet state $D_s$ ($^1$P$_1$) is located at about 55 MeV
above the $\Lambda p$ threshold, and the triplet states
$D_t$ ($^3$P$_{0,1,2}$) are located at about 95 MeV above the
$\Lambda p$ threshold.
This corresponds for the lowest-lying dibaryon $D_s$ 
to an invariant mass of  2109~MeV which  is located
between the $\Lambda $-proton and $\Sigma $-nucleon thresholds.
The resonance $D_s$ is a
singlet state of a
four quark-two quark cluster configuration
$({\rm Q}^4)_3\bigotimes({\rm Q}^2)_{3^*}$
with angular momentum $L=1$,
spin $S=0$, total angular momentum and parity $J^{P}=1^-$. The isospin is $I=1/2$.
The only particle-decay channel is the $^1{\rm P}_1$ wave of the $\Lambda {p}$ system.
This decay is hindered by the relative $P$-wave centrifugal
barrier between the two clusters and
the stability of the $({\rm Q}^4)_3\bigotimes({\rm Q}^2)_{3^*}$
configuration against dissociation into two color singlet (${\rm Q}^3$) clusters.
In other words, the $\Lambda {p}$ channel represents only a small
piece  of the $D_s$ wave function.
As a consequence, the total width $\Gamma$ of $D_s$
is predicted to be less than about 1~MeV \cite{aer84,aer85}.
In addition, the predicted width  depends on the resonance mass $M_r$, see Fig. 2 of \cite{aer84}.
We note that there is a rather large uncertainty of the predicted resonance masses $M_r$.
Therefore, the production of the dibaryon $D_s$ has been
studied theoretically in the  invariant mass range between 2.06 and 2.10 GeV/c$^2$ \cite{aer85}.
In the present paper, we consider the invariant mass range between 2.058 and 2.105 GeV/c$^2$
which corresponds to resonance energies $E_r$ between 4 MeV and 51 MeV.


We deduce from Fig.~2 of \cite{aer84} $\Gamma = 26$~keV for
$M_r=2109$~MeV ($E_r=55$ MeV).
Extrapolating to $M_r=2096.5$~MeV ($E_r=42.5$~MeV)
yields $\Gamma=15$~keV.
We note that these total  widths are
lower limits. 
The decay width calculation involves an integration of the dibaryon wave function
over the radial distance $r$. Such integrals are dominated by the
contributions from  small radii $r$ which give a lower limit of $\Gamma$.
Including the larger radii gives an upper limit.
Then, the lower limit has to be multiplied by 64/3 to obtain an upper limit.
In addition, the width increases by about a factor of 2 if
Hulth\`en rather then oscillator wave functions are used.
Thus, an
upper limit is obtained
if the lower limit is multiplied by a factor 128/3 \cite{aer84}.
We get the following ranges for the predicted total width:
$\Gamma=[15,640]$~keV for $M_r=2096.5$~MeV ($E_r=42.5$~MeV) and
$\Gamma=[26,1109]$~keV for $M_r=2109$~MeV ($E_r=55$~MeV).
As one can see, the predicted widths depend strongly on the invariant mass of the dibaryon.

We emphasize that a possible resonance $D_s$
below the $\Sigma^+ {n}$ threshold (2128.935 MeV)
can only decay
into $\Lambda +p$. Therefore, the elastic width $\Gamma_{el}$
can be assumed to be equal to the total width $\Gamma$, i.e. 
$\Gamma_{el}/\Gamma=1$.

\section{Simulation of a narrow resonance in the total cross section of $\Lambda p \to \Lambda  p$ scattering}

We first discuss the effect of a single isolated narrow resonance in
the total cross section of the $\Lambda + p \to \Lambda + p$ scattering
and we compare it with existing data from bubble chamber experiments
\cite{gro63,pie64,fhint:ale68,fhint:sec68,kad69,hau77}.
We start with the ansatz
\begin{equation}
\sigma_{tot}=\sigma_{tot}^{nr}+\sigma_{tot}^r.
\end{equation}
Here, $\sigma_{tot}^{nr}$ is the nonresonant part and $\sigma_{tot}^r$ the resonant part
of the total cross section.
Assuming S-waves and
taking the effective range approximation
the nonresonant part of the total  cross section may be written \cite{gold64}
\begin{equation}
\sigma_{tot}^{nr}=\frac{1}{4} \frac{4\pi}{k^2+\left(-\frac{1}{a_s}+\frac{r_sk^2}{2}\right)^2}
+\frac{3}{4}\frac{4\pi}{k^2+\left(-\frac{1}{a_t}+\frac{r_tk^2}{2}\right)^2}
\label{sigtot}
\end{equation}
Here, $a_s$ and $a_t$ are the singlet and triplet scattering lengths,
$r_s$ and $r_t$ are the singlet and triplet effective ranges and
$k=q/\hbar$ is the
wave number corresponding to the c.m. momentum $q$.
We take the effective range parameters $a_s=-2.43$~fm, $r_s=2.21$~fm,
$a_t=-1.56$~fm, $r_t=3.7$~fm as determined in \cite{bud10}.
\begin{figure}[t!]
\begin{center}
\includegraphics[width=0.5\textwidth]{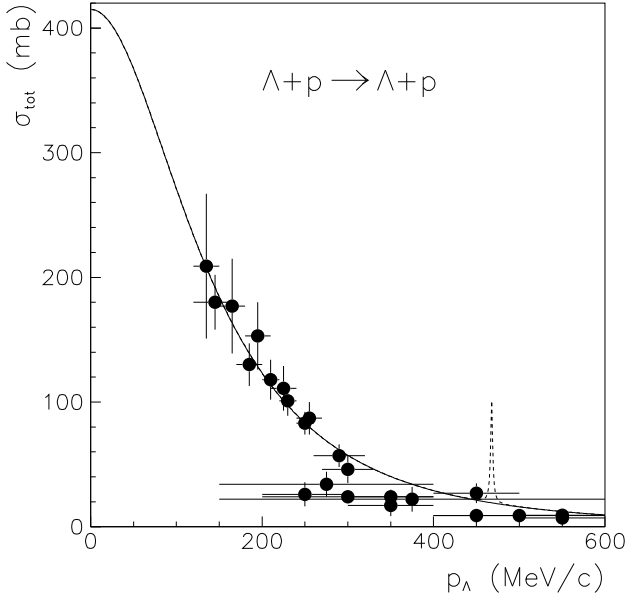}
\end{center}
\caption{Total $\Lambda p \to \Lambda p$ cross section  vs laboratory momentum $p_{\Lambda}$.
Solid line: effective range approximation of $\sigma_{tot}^{nr}$.
Dashed line: Simulation of a resonance signal $\sigma_{tot}^r$
without folding with the effective resolution function
for a resonance in the $^1P_1$ channel with
$E_r=42.5$~MeV and $\Gamma=500$~keV.
The data are from \cite{gro63,pie64,fhint:ale68,fhint:sec68,kad69,hau77}}
\label{sigtot_f}
\end{figure}

The resonant part of the total cross section can  be approximated by the
Breit-Wigner resonance formula,
\begin{equation}
\sigma_{tot}^r = \frac{2J_r+1}{(2s_1+1)(2s_2+1)}\frac{\pi}{k^2}\frac{\Gamma^2}{(E-E_r)^2+(\Gamma/2)^2}.
\label{sigres}
\end{equation}
Here, $E$ is  the total kinetic energy in the  c.m. system,
$k=q/\hbar$ is the wave number corresponding to the c.m. momentum $q$
and we assume $\Gamma_{el}=\Gamma$.
We take the narrow resonance $D_s$ as an example. The resonance is predicted to occur in
the partial wave $^1P_1$. Therefore, we assume $J_r=1$.
As resonance energy we take
 $E_r=42.5$~MeV corresponding to $\sqrt{s}=2096.5$~MeV of the peak seen by 
Siebert et al. \cite{fhint:sie94}.
At this energy, the nonresonant potential scattering is dominated
by the partial waves $^1S_0$ and $^3S_1$ and the nonresonant contributions of the $P$ waves can be neglected
in Eq.~(\ref{sigtot}).
The same holds true for a possible interference between the resonant and nonresonant $^1P_1$ amplitude
in Eq.~(\ref{sigres}).  
The J\"ulich hyperon-nucleon model \cite{haid05} yields for instance phase shift predictions
with $\delta<5^{\circ}$ for $^1P_1$ and $^3P_{0,1,2}$.

In Fig.~\ref{sigtot_f} we show the effect of the $D_s$ resonance
with $J_r=1$ and $E_r=42.5$~MeV for
$\Gamma=500$~keV
without taking the effect of a finite energy resolution into account.
We take $\Gamma=500$~keV as an example. A resonance  with
$\Gamma=15$~keV would appear as a sharp needle in Fig.~\ref{sigtot_f}.
The maximum  of the resonance signal  at $E=E_r$ is very large,
$\sigma_{tot}^{r}(E_r)/\sigma_{tot}^{nr}(E_r)=4.2$.
Such a resonance could be easily observed provided the effective energy resolution
would be sufficiently high. Unfortunately, the bubble chamber experiments suffer from
low-energy resolution as well as from low statistical accuracy.
For instance data points 
near the resonance exhibit 
bin widths of 50~MeV/$c$ ($\Delta E_{cm}=8.4$~MeV) 
and 100~MeV/$c$ ($\Delta E_{cm}=16.8$~MeV).
These  bin widths
are so large that the effect of a narrow resonance is completely diluted.

\section{Narrow resonance embedded in the continuum of
$pp \to K^+ (\Lambda p)$}

The reaction  $pp \to K^+ (\Lambda p)$ can be described by factorizing the
reaction amplitude in terms of a production amplitude
and a final state enhancement factor.
The method of parametrizing the FSI enhancement factor by the inverse Jost function \cite{gold64}
is described in \cite{bud10,hint04}.
Taking the spin statistical weights into account
the nonresonant double differential cross section  may be written as
\begin{equation}
\frac{{\rm d}^2\sigma^{nr}}{{\rm d}\Omega_{K}{\rm d}M_{\Lambda{p}}}= \Phi_3 \left[\, \frac{1}{4} \,
 |{M}_s|^2 \, \frac{q^2+\beta_s^2}{q^2+\alpha_s^2}
+\, \frac{3}{4} \, |{M}_t|^2 \,
\frac{q^2+\beta_t^2}{q^2+\alpha_t^2}\right].
\label{miss}
\end{equation}
Here, $|{M}_s|^2$ and $|{M}_t|^2$ are the singlet and triplet production
matrix elements squared, $q$ the internal c.m.-momentum of the $\Lambda p$ system,
 $\alpha_s$, $\beta_s$, $\alpha_t$, $\beta_t$ the singlet and triplet potential parameters,
and $\Phi_3$ the ratio of the three-body phase space distribution
and the incident flux factor.
The potential parameters $\alpha$ and $\beta$ can be used to establish
phase-equivalent Bargmann potentials~\cite{bar49b}.
They are related
to the scattering lengths $a$, and effective ranges $r$ of the low-energy
$S$ wave scattering,
$\alpha=(1-\sqrt{1-2r/a})/r$,
$\beta=(1+\sqrt{1-2r/a})/r$.
Thus, the shape of the missing-mass spectrum
(\ref{miss}) depends on $|M_s|^2$, $|M_t|^2$ and the
effective range parameters $a_s$,
$r_s$, $a_t$ and $r_t$.
These parameters have been deduced by a combined fit of the $pp \to K^+ (\Lambda p)$
data and the total $\Lambda {p}$ cross section data \cite{bud10}.
In principle one could equally well take a fit with three spin-averaged parameters
$|\bar{M}|^2$, $\bar{a}$ and $\bar{r}$  (see \cite{bud10}) 
or another three-parameter description of the missing-mass spectrum as in e.g. 
\cite{gasp04}.

A narrow resonance $D_s$ embedded in the $\Lambda p$  continuum of the
reaction $pp \to K^+ (\Lambda p)$ can be described
using the following equation
\begin{gather}
\frac{{\rm d}^2\sigma^{r}}{{\rm d}\Omega_{K}{\rm d}M_{\Lambda{p}}}=
\frac{{\rm d}\sigma^{r}}{{\rm d}\Omega_K}
\frac{2 M_{\Lambda {p}}}{\pi}\frac{M_r \Gamma }{(M^2_{\Lambda {p}} - M^2_r)^2 + M_r^2 \Gamma^2}.
\label{resonance-1}
\end{gather}
This equation corresponds to a formula derived by Pilkuhn \cite{pil79}
in order to describe  the reaction $a+b \to c + d$ where $d$ is a narrow resonance
decaying into particles 1 and 2, $d\to {1} + {2}$.
Here, ${\rm d}\sigma^{r}/{\rm d}\Omega_K$ represents the differential cross section 
for the production of the resonance $D_s$, $M_r$ the resonance mass, $\Gamma$ the total width
and $M_{\Lambda {p}}$ the invariant mass of the $\Lambda {p}$ system. 
The relativistic Breit-Wigner form in (\ref{resonance-1}) can be approximated by
the corresponding nonrelativistic Breit-Wigner form,
\begin{gather}
\frac{{\rm d}^2\sigma^{r}}{{\rm d}\Omega_{K}{\rm d}M_{\Lambda{p}}}=
\frac{{\rm d}\sigma^{r}}{{\rm d}\Omega_K}
\frac{1}{2\pi}\frac{\Gamma }{(M_{\Lambda {p}} - M_r)^2 + (\Gamma/2)^2}.
\label{resonance-2}
\end{gather}
In passing, we note that the integral over the Breit-Wigner distribution is normalized to one,
\begin{equation}
\int_{-\infty}^{\infty} \frac{1}{2\pi} \frac{\Gamma }{(M_{\Lambda {p}} - M_r)^2 + (\Gamma/2)^2}
{\rm d}M_{\Lambda {p}}=1.
\end{equation}
The double differential cross section including a possible narrow resonance $D_s$ in the
$\Lambda {p}$ system may be written by combining (\ref{miss}) and (\ref{resonance-1}),
\begin{equation}
\frac{{\rm d}^2\sigma}{{\rm d}\Omega_{K}{\rm d}M_{\Lambda{p}}}=
\frac{{\rm d}^2\sigma^{nr}}{{\rm d}\Omega_{K}{\rm d}M_{\Lambda{p}}}+
\frac{{\rm d}^2\sigma^{r}}{{\rm d}\Omega_{K}{\rm d}M_{\Lambda{p}}}.
\label{ansatz}
\end{equation}
We mention again that possible $P$ wave contributions in the $\Lambda{p}$ interaction
are negligibly small at low $\Lambda{p}$ c.m. energies \cite{haid05}. Therefore,
the effect of an interference with the nonresonant $^1P_1$ decay channel can be neglected.
The relative contribution of a narrow resonance $D_s$ to the double differential cross section
depends on the production cross section
${\rm d}\sigma^{r}/{\rm d}\Omega_K$ for $(pp\to K^+ D_s)$.
In the present paper, we deduce  upper limits for the production cross section.

\begin{figure}[t!]
\begin{center}
\includegraphics[width=0.5\textwidth]{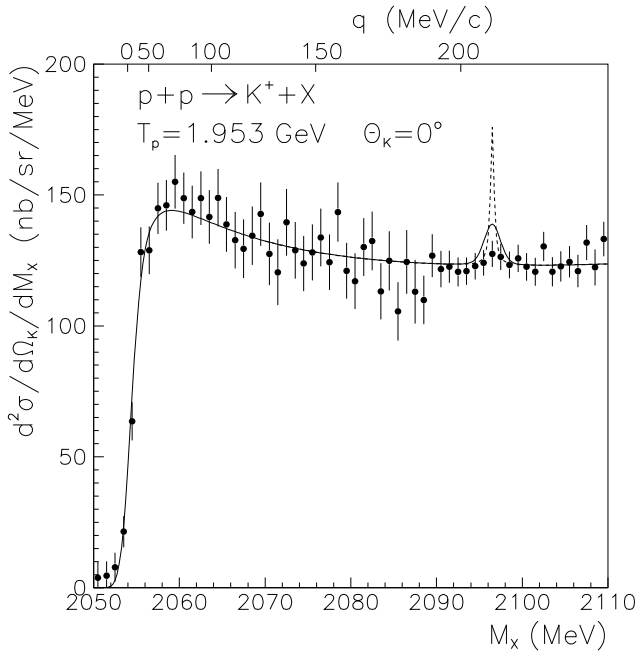}
\caption{
Missing-mass spectrum of the reaction $p+p \to K^+ + X$ with $X=(\Lambda p)$
measured at $T_p=1.953$~GeV and $\Theta_K=0^{\circ}$.
The upper axis indicates
the c.m. momentum $q$ of the $\Lambda p$ system.
Solid line: FSI fit curve with
resonance signal excluded by the $\chi^2$ test.
Dashed line: Same resonance signal without folding with the effective resolution function.
Resonance parameters:
${\rm d}\sigma^r/{\rm d}\Omega_K=42$~nb/sr, $M_r=2096.5$~MeV, $\Gamma=500$~keV.
}
\label{spectrum}
\end{center}
\end{figure}

\section{Upper limits for a narrow resonance}

A narrow $S=-1$ resonance $D_s$ as predicted by \cite{aer84,aer85} is not visible in the data
(see Fig.~\ref{spectrum}). There may be two reasons for it.
(i) The predicted narrow   $S=-1$ resonance $D_s$ does not exist at all
in the invariant mass region below 2110~MeV.
(ii) The production cross section
${\rm d}\sigma^{r}/{\rm d}\Omega_K(pp\to K^+ D_s)$
is too small.
In the following, we deduce upper limits for the production cross section
in the given invariant mass range.

\subsection{Effective resolution function}

We note that the theoretical expression for the double differential cross section
(\ref{ansatz}) must be folded with the effective resolution function
$f(M,M')$ before comparing with the data.
The effective resolution function is represented by a Gaussian density distribution
with a standard deviation of $\sigma_M=0.84$ MeV. 
It was deduced in \cite{bud10}  by a least-square fit to 
the sharply rising
double differential cross section
${\rm d}^2\sigma/({\rm d}\Omega_{K}{\rm d}M_{\Lambda{p}})$
near the $(\Lambda {p})$ threshold, see Fig.~\ref{spectrum}.
As discussed in Sec.~\ref{experiment} the effective resolution function
is constant in the full missing-mass range between  2050~MeV and 2110~MeV.
\begin{figure}[t!]
\begin{center}
\includegraphics[width=0.5\textwidth]{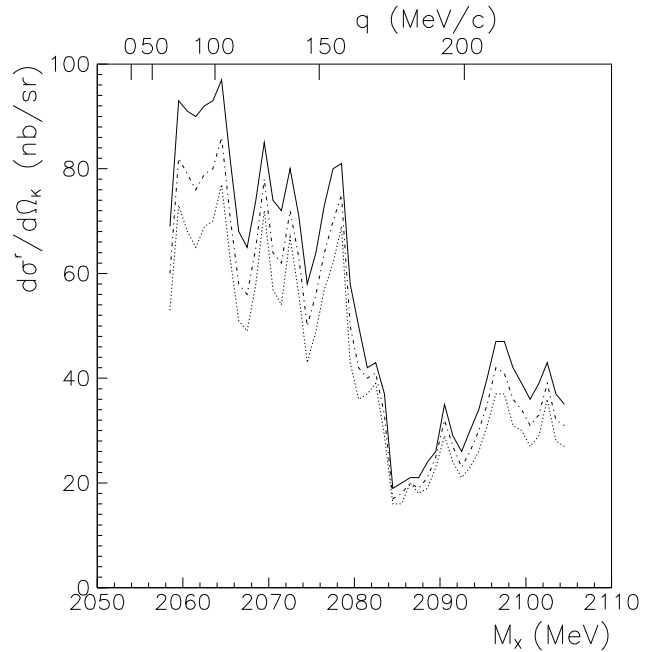}
\caption{
Upper limits with 99\% confidence level
of the production cross section
${\rm d}\sigma^r/{\rm d}\Omega_K$ (nb/sr)
for $\Gamma=100$ keV (dotted line), $\Gamma=500$ keV (dashed-dotted line), and
$\Gamma=1.0$ MeV (solid line).
}
\label{upper-limit}
\end{center}
\end{figure}

\subsection{$\chi^2$ test}

The compatibility of the experimental data with the hypothesis of a narrow
resonance in the invariant mass region below 2110~MeV
has been studied  using  a $\chi^2$ test. 
Using this method we determine the upper limits of
the production cross section
${\rm d}\sigma^{r}/{\rm d}\Omega_K(pp\to K^+ D_s)$
in the $\Lambda p$ invariant mass range between 2058 and 2105 MeV,
see Fig.~\ref{upper-limit}.
We only include the data measured within 
$\pm 4$~MeV
around the assumed resonance mass $M_r$. The data are binned in 1 MeV wide bins;
thus nine data points are used 
for the $\chi^2$ test. Including more data would  dilute
the information, since an
excursion with a width of about $\sigma_M = 0.84$~MeV is negligibly small for 
$|M_{\Lambda {p}}-M_r|\geq 5$~MeV, see Fig.~\ref{spectrum}.


The $\chi^2$ obtained from a comparison of the hypothesis to the data
is subject to a standard $\chi^2$ test \cite{pdg10}.
We assume that the $\chi^2$ is statistically distributed according to
a $\chi^2$ probability density function $f(\chi^2,n_d)$ with the appropriate number of
degrees of freedom, $n_d$. Thus, we can calculate the confidence level (CL)
with which we can support or falsify the hypothesis.
If the hypothesis is true, the probability $P$ to obtain a larger  $\chi^2$
than the observed one in an infinite repetition of the experiment
is given by
\begin{equation}
P=\int_{\chi^2}^{\infty}f(z,n_d) {\rm d}z.
\end{equation}
If $P$ is very small, either the hypothesis is wrong or the current
measurement suffers from a very unlikely statistical fluctuation.
The CL for {\em excluding} the hypothesis of a narrow resonance with a certain
production cross section is given by $1-P$.
Thus, we have $P=1$~\% for a confidence level of 99~\%, i.e. the probability to observe a larger
$\chi^2$ when repeating the experiment is only 1~\%.

We use the number of data points within the invariant mass interval $M_r\pm 4$~MeV
as the number of degrees of freedom, $n_d$.
Thus, we neglect the degrees of freedom introduced by fitting
the $\Lambda p$ missing-mass spectrum. Three fit parameters, i.e. the spin-averaged parameters
$\bar{M}^2$, $\bar{a}$ and $\bar{r}$, are sufficient in order to reproduce the
measured missing-mass spectrum. They cannot be considered as free parameters in the present
hypothesis test. We
mention that there is no significant effect on the deduced upper limits
due to the errors 
of those parameters.
Thus, we have  $n_d=9$ and $\chi^2/n_d=2.41$ for $P=1$~\%.


\subsection{Upper limits for the production cross section
${\rm d}\sigma^{r}/{\rm d}\Omega_K(pp\to K^+ D_s)$}

The upper limits for the production cross section ${\rm d}\sigma^{r}/{\rm d}\Omega_K(pp\to K^+ D_s)$
are shown in Fig.~\ref{upper-limit} for $\Gamma=100$ keV (dotted line), $\Gamma=500$ keV (dashed-dotted line), and
$\Gamma=1.0$ MeV (solid line). We note that the upper limits depend slightly on the total width $\Gamma$.
For very small total widths, i.e. $\Gamma\leq 100$ keV
one can take the upper limits for $\Gamma=100$~keV.
The differences between the upper limits for $\Gamma<100$ keV and $\Gamma=100$ keV are negligible.

The upper limits exhibit rather strong fluctuations due to the statistical fluctuations of the data.
In the lower part of the missing-mass spectrum between 2058 and 2080 MeV,
the upper limit varies between
43 and 77 nb/sr for $\Gamma=100$ keV, 50 and 86 nb/sr for $\Gamma=500$ keV, and
50 and 97 nb/sr for $\Gamma=1.0$ MeV.
In the region between 2080 and 2105 MeV the upper limits are roughly a factor of 2
smaller. This is 
because four data points between 2080 and 2090 MeV are below the fit curve 
and that 
the missing-mass region between 2090 and 2105 MeV has been measured 
with a much higher statistical accuracy.
There, the upper limit varies between
21 and 37 nb/sr for $\Gamma=100$ keV, 23 and 42 nb/sr for $\Gamma=500$ keV and
26 and 47 nb/sr for $\Gamma=1.0$ MeV.


\section{Conclusion and Discussion}

The effect of a single isolated narrow resonance in
the total cross section of the free $\Lambda {p}$ scattering
has been studied (see Fig.~\ref{sigtot_f}).
A narrow strangeness $S=-1$ resonance in the $^1P_1$ channel
is predicted to be characterized by a rather large resonance signal.
Therefore, the free $\Lambda {p}$ scattering would be ideal
to search for a narrow resonance.
However, the effective resolution and the statistical accuracy
of existing $\Lambda {p}$ scattering data are too low for
a stringent upper limit search.

The reaction $pp\to K^+ + (\Lambda p)$ has been measured at $T_p=1.953$ GeV and
$\Theta=0^{\circ}$ with a high missing-mass resolution.
A narrow strangeness $S=-1$ resonance $D_s$ as predicted by \cite{aer84,aer85} is not visible in the data
(see Fig.~\ref{spectrum}). 
The missing-mass range between 2090 and 2110 MeV
has been studied with an especially high statistical accuracy.
The apparent small enhancement  near 2096.5 MeV 
(see Fig. \ref{spectrum}) corresponds
to an excursion of only 1.5 standard deviations.
Upper limits of the production cross section
${\rm d}\sigma^{r}/{\rm d}\Omega_K(pp\to K^+ D_s)$
are deduced for resonance energies between  2058 and 2105 MeV, see Fig.~\ref{upper-limit}.
In the missing-mass region between 2090 and 2105 MeV the upper limit varies between
21 and 37 nb/sr for $\Gamma=100$ keV, 23 and 42 nb/sr for $\Gamma=500$ keV and
26 and 47 nb/sr for $\Gamma=1.0$ MeV.

Unfortunately, there are no theoretical predictions of the production cross section
${\rm d}\sigma^{r}/{\rm d}\Omega_K(pp\to K^+ D_s)$.
Aerts and Dover gave estimates of production cross sections
of strangeness $S=-1$ dibaryon states $D_{s,t}$
for the reactions $K^- + d \to \pi^- +D_t$ and $K^- + ^3He \to \pi^+ + n + D_s$ \cite{aer85}. 
Differential cross sections of order 1~$\mu$b/sr were predicted.

The transition  $pp \to K^+ D_s$ from the ${p}{p}$ entrance channel requires orbital
angular momentum transfer $\Delta l=0$ and 2.
Thus, the production cross section
${\rm d}\sigma^{r}/{\rm d}\Omega_K(pp\to K^+ D_s)$ should peak at the scattering angle $\Theta=0^{\circ}$.
The  reaction $pp \to K^+ (\Lambda {p})$ has been measured at $\Theta=0^{\circ}$,
i.e. at the maximum of the angular distribution. 
Therefore, one expects maximum sensitivity when searching for a narrow strangeness $S=-1$ resonance $D_s$
with the reaction $pp \to K^+ D_s$.

It is interesting to note that the upper limit of the production cross section 
${\rm d}\sigma^{r}/{\rm d}\Omega_K(pp\to K^+ D_s)$ is nearly independent of the 
total width $\Gamma$ as long as $\Gamma < \sigma_M$. This is due to the fact
that the integral over the Breit-Wigner distribution (\ref{resonance-2}) is constant (equal to one),
that means it does not depend on $\Gamma$.
Thus, folding
a narrow Breit-Wigner distribution (\ref{resonance-2})   with a wider  Gaussian distribution
yields a resonance signal which depends only on the production cross section 
and the effective resolution width $\sigma_M$ but not on $\Gamma$.
This is a great advantage when searching for an extremely narrow resonance.
The sensitivity in the search for a narrow resonance can be increased substantially by
increasing the missing-mass resolution, i.e.
by decreasing the resolution width $\sigma_M$.
Another less efficient  way is to increase the statistical and systematic accuracy.

In contrast to previous experiments, the present experiment exhibits a rather high missing-mass resolution.
In previous experiments, the missing-mass resolution has been very much lower.
Hogan et al. \cite{fhint:hog68} studied the reaction $pp \to K^+ (\Lambda {p})$ at 
bombarding energies between 2.5 and 3.0 GeV and scattering angles $20^{\circ}$, $30^{\circ}$ and $40^{\circ}$.
The momentum resolution of the detected kaons ($\Delta {p}/{p}$) 
ranged from 1.5~\% at $20^{\circ}$ to 3~\% at $40^{\circ}$.
The number of data points per MeV/$c$ and the effective missing-mass resolution was so low that they were not able to see
the marked FSI enhancement near the $\Lambda {p}$ threshold.
Reed et al. \cite{fhint:ree68} studied the reaction $pp \to K^+ (\Lambda {p})$ at 
2.40 and 2.85 GeV and $0^{\circ}$,
$17^{\circ}$ and $32^{\circ}$.
The momentum resolution of the outgoing kaons at $0^{\circ}$ corresponded to $\Delta {p}/{p}=1.5$~\%.
This resolution was also too low to search for a narrow resonance with $\Gamma\leq 1$~MeV.
The same holds true for the exclusive measurements of the reaction $pp \to K^+ \Lambda {p}$
by the COSY-TOF Collaboration \cite{bil98,abd06,abd10}.

As mentioned in the introduction, Siebert et al. \cite{fhint:sie94} performed the first high-resolution
study of the reaction $pp \to K^+ (\Lambda {p})$. 
Sharp peaks in the missing-mass spectra 
seen there
have not been confirmed by the present experiment.
In this context we note that the reaction  $pp \to K^+ (\Lambda {p})$ has been studied 
\cite{fhint:sie94} at $T_p=2300$~MeV  at four scattering angles,
6$^{\circ}$, 8.3$^{\circ}$, 10.3$^{\circ}$, and 12$^{\circ}$.
The peak near $2096.5\pm 1.5$ appeared only 
in the missing-mass spectrum measured
at $10.3^{\circ}$  but not
at 6$^{\circ}$, 8.3$^{\circ}$ and 12$^{\circ}$.  
At $T_p=2700$~MeV, the interesting missing-mass range  has been studied at two 
scattering angles, 12.6$^{\circ}$ and 20$^{\circ}$.
The peak near $2098.0\pm 1.5$ appeared only at $\Theta_K=12.6^{\circ}$ but not
at 20$^{\circ}$.
We therefore conclude that the peaks near $2096.5\pm 1.5$ and $2098.0\pm 1.5$~MeV  
must be attributed to a statistical fluctuation of the nonresonant cross section.

Summarizing, the reaction $pp\to K^+ (\Lambda {p})$ has been measured with
a high missing-mass resolution at $T_p=1.953$~GeV and $\Theta=0^{\circ}$.
A narrow strangeness $S=-1$ resonance $D_s$ 
is not visible in the missing-mass spectrum. Upper limits for the 
production cross section of $pp\to K^+ {D_s}$ have been deduced.

\section*{ACKNOWLEDGEMENT}  
We acknowledge helpful discussions with J. Haidenbauer and C. Hanhart.

\end{document}